\documentclass[fleqn,twoside]{article}
\usepackage[headings]{espcrc2}

\readRCS
$Id: espcrc2.tex,v 1.2 2004/02/24 11:22:11 spepping Exp $
\usepackage{graphicx}
\usepackage[figuresright]{rotating}

\newcommand{\AmS}{{\protect\the\textfont2
  A\kern-.1667em\lower.5ex\hbox{M}\kern-.125emS}}

\title{Semi-analytic calculation of the monopole order parameter in QCD}

\author{Adriano Di Giacomo}
\address{Dipartimento di Fisica Universita' Pisa and INFN Sezione di Pisa, 
         \\ 
        3 Largo B. Pontecorvo 56127 Pisa ,ITALY}
               
\runtitle{Semi-analytic calculation..}
\runauthor{A.Di Giacomo}

\begin{document}

\begin{abstract}
The monopole order parameter of QCD is computed in terms of gauge invariant field strength correlators. Both quantities are partially known from  numerical simulations on the lattice. A new insight results on the structure of the confining vacuum.
\vspace{1pc}
\end{abstract}

\maketitle

\section{Introduction}
The mechanism of confinement by dual supercoductivity of the $QCD$ vacuum \cite{1}\cite{2}\cite{3}
is confirmed by numerical simulations on a lattice.
 Chromoelectric flux tubes  connecting $q-\bar q$ pairs produced by dual Meissner effect are indeed
 observed in lattice configurations with the expected form of the electric and magnetic fields\cite{3'}.
 An extensive analysis has been performed by exploring the vacuum by means of an order parameter
 $\langle \mu \rangle$ \cite{4,ddpp,dp,5,6} which is the vacuum expectation value of a magnetically charged operator $\mu$. In the confined phase  $\langle \mu \rangle\neq 0$ , which implies Higgs breaking of the magnetic U(1) symmetry, in the deconfined phase  $\langle \mu \rangle=0$ and the magnetic charge is superselected.
 In SU(N) gauge theory there exist $N-1$ magnetic charges and $N-1$ independent operators $\mu^a , (a=1, ..,N-1)$, which create monopoles of the species $a$\cite{16}. They can be written
  \begin{equation}
 \mu^a (\vec x, t)=e^ { {iq\over {2g}}\int d^3y \vec b_{\perp}(\vec x-\vec y)Tr(\Phi^a \vec E) (\vec y, t)}
  \end{equation}
  $q$ is an integer $\vec b_{\perp}$  is the field of a Dirac monopole with $\vec \nabla\vec b_{\perp}=0$ and $\vec \nabla\wedge \vec b_{\perp} = {\vec r\over r^3} + Dirac - string$.
  \begin{equation}
  \Phi^a(x) \equiv U(x,y) \Phi^a_{d} U^{\dagger}(x,y) 
 \end{equation} 
   with   $U(x,y)$ an arbitrary gauge transformation. We will take for U a parallel transport to $x$ from a reference point $y$ along
  a path $C$ .
  
   $\langle \mu^a \rangle$ is gauge invariant. In Eq(2) 
   \begin{eqnarray}
                                                   <--a -->&.&<-(N-a)>   \nonumber     \\  
  \Phi^a_{d} \equiv diag({{N-a}\over N},.,{{N-a}\over N}&,& -{a\over N},.,-{a\over N})
  \end{eqnarray}
  In the gauge in which $\Phi^a =\Phi^a_d$ (Abelian Projection) 
  \begin{equation}
   \mu^a (\vec x, t)=e^ { {iq\over {2g}}\int d^3y \vec b_{\perp}(\vec x-\vec y)\vec E^a_{\perp} (\vec y, t)}
   \end{equation}
   where  $\vec E^a_{\perp}$ is the transverse 
    chromoelectric field along the color direction $T^a$
   \begin{eqnarray}
            a&.&a+1 \nonumber \\
   T^a= diag(0,...0,1&,&-1,0,..0)
   \end{eqnarray}
   $\vec {E^a}_{\perp}$ is the conjugate momentum to $\vec {A^a}_{\perp}$. Hence
\begin{equation}
   \mu^a(\vec x,t) |\vec A^a_{\perp}(\vec y,t)\rangle= |\vec A^a_{\perp}(\vec y,t)+ {q\over {2g}}\vec b_{\perp}(\vec x-\vec y)\rangle
   \end{equation}
    $\mu^a (\vec x, t) $ creates a Dirac monopole at   $(\vec x, t)$ in the residual gauge symmetry
    after abelian projection.
     It proves convenient  to use instead of    $\langle \mu^a \rangle$ the susceptibility 
   \begin{equation}
   \rho^a \equiv {\partial \over{\partial \beta}}  ln\langle \mu^a \rangle
   \end{equation}
   Here $\beta$ is the usual variable of the lattice formulation $\beta\equiv {{2N}\over g^2}$.
 Since at $\beta=0$ $\mu^a = 1$, 
\begin{equation}
    \langle \mu^a \rangle = e^{\int ^{\beta}_0 d{\beta} \rho^a(\beta)}
    \end{equation}
    $\rho^a$ has been measured by lattice simulation for various gauge theories:compact $U(1)$ \cite{dp},$SU(2)$ \cite{5} , $SU(3)$ \cite{6} and   $N_f=2$ $QCD$ \cite{17} .
    In all these systems $\rho^a \to finite $ in the confined phase in the thermodynamical limit $V\equiv L_s^3 \to \infty$.
   By use of Eq(8) this means   $\langle \mu^a \rangle \neq 0$ .
   
   In the deconfined phase  $T > T_c$
   
    \begin{equation}
    \rho^a \approx -|c| L_s+c' $  or  $\langle \mu^a \rangle = 0
    \end{equation}
    or, again by Eq(8)   $\langle \mu^a \rangle = 0$ .
    
    In the critical region   $T\approx T_c$ the scaling law holds
    \begin{equation}
      {\rho^a \over L_s^{1\over \nu}}\approx f(\tau L_s^{1\over \nu})
      \end{equation}
   Here  $\tau \equiv  1-{T\over T_c} $ ,  $\nu$  is the critical index of the correlation length of the order parameter.
     $\rho^a$ is independent of the choice of the abelian projection \cite{7} \cite{8}  \cite{71}.
    Expanding the exponential which defines $\langle \mu^a \rangle $ one has
    \begin{eqnarray}
 \langle \mu^a \rangle = \Sigma_0^{\infty} ({{iq}\over{2N}})^n {1\over {n!}} \int d^3y_1 ..d^3 y_n 
 b^{i_1}(\vec x -\vec y_1)..\nonumber
 \\ b^{i_n}(\vec x -\vec y_n) \langle (\Phi^a.\vec E)_{{i_1},{\vec y_1}} (\Phi^a.\vec  E)_{{i_n},{\vec y_n}}\rangle
 \end{eqnarray}
 The  notation is $(\Phi^a.\vec E)_{i,x} \equiv Tr[\Phi^a (x). E^i (x)$.
 
 The  $vev's$ in Eq(11) are gauge invariant field strength correlators.
 
We shall identify these correlators with those of the Stochastic Vacuum approach to QCD \cite{13a} \cite{13b} \cite{13}.
    \section{Cluster Expansion of $\langle \mu^a \rangle $}
In $QCD$ all observables can be expressed in terms of gauge invariant field strength correlators.
 The basic idea of the stochastic vacuum approach is to perform a systematic cluster expansion of
 the correlators and truncate it typically retaining only the clusters up to order two.
 Since the one point cluster $\langle (\Phi^a.\vec E)\rangle  = 0$, only two point clusters will survive
 the truncation, and hence only even terms in the expansion Eq(11) , which will be approximated as products of two point correlators  $\Phi^a_{i_1,i_2}(\vec y_1 -\vec y_2)= \langle (\Phi^a.\vec E)_{{i_1},{\vec y_1}} (\Phi^a.\vec  E)_{{i_2},{\vec y_2}}\rangle $.
  There is a  combinatorial factor $(2n - 1)!!$ for  the term of order $2n$   .
 Eq(11) becomes
 \begin{equation}
  \langle \mu^a \rangle = e^{- {{q^2}\over {8g^2}} \int d^3 y_1 \int d^3 y_2 \Phi^a_{i_1,i_2}(\vec y_1 -\vec y_2)b^{i_1}_{\perp}(\vec y_1) b^{i_2}_{\perp}(\vec y_2)}
  \end{equation}

   or, since  $\beta = {{2N}\over{g^2}}$ 
    \begin{eqnarray}
  \rho^a=- {{q^2}\over {16N}}{\partial \over \partial \beta}[\beta \int d^3 y_1 \int d^3 y_2 \nonumber \\ \Phi^a_{i_1,i_2}(\vec y_1 -\vec y_2)b^{i_1}_{\perp}(\vec y_1) b^{i_2}_{\perp}(\vec y_2)]
  \end{eqnarray}
   We shall identify $\Phi^a_{i_1,i_2}$ with the two point correlators defined with a straight line parallel transport. These correlators  are measured on the lattice \cite{10}\cite{11}\cite{12} , and are used as an input in  stochastic  $QCD$.   For those correlators $\langle E^a E^b \rangle = \delta^{ab} \Phi$ , so that $\rho^a$ is independent on $a$ . This is also the case in  lattice determinations of $\rho^a$ \cite{6}.
   
  The  the cluster expansion is generically expected to work at large distances, and in the study of confinement we are looking for infrared properties . Anyhow a direct check of it can be obtained by looking at the dependence of $\rho^a$ on  $q$. The truncated $\rho^a$ is proportional to $q^2$ : higher correlators would introduce terms proportional to higher powers of $q$. Old data \cite{5} \cite{9'} seem to agree with   $q^2$ but a  systematic study  of this dependence will be done.

 \section{The Field Correlators}

 A general parametrization of field strength correlators dictated by invariance arguments\cite{13a} \cite{13b} is 
 \begin{eqnarray}
 \Phi^{ab}_{\mu_1,\nu_1,\mu_2\nu_2}(z_1-z_2) \equiv  {1\over N}\langle Tr F^a_{\mu_1 \nu_1}(z_1)   \nonumber \\ V(z_1,z_2) F_{\mu_2,\nu_2}(z_2)V^{\dagger}(z_1,z_2)\rangle
 \end{eqnarray}
 \begin{eqnarray}
 \Phi^{ab}_{\mu_1,\nu_1,\mu_2\nu_2}(z_1-z_2) = \delta^{ab}   \nonumber  \\    (   D(z_1-z_2)  [\delta_{\mu_1\mu_2 } \delta_{\nu_1\nu_2 }-\delta_{\mu_1\nu_2 }\delta_{\nu_1\mu_2 }] \nonumber \\+{1\over 2}{\partial\over{ \partial z_{\mu_1}}}[D_1(z_1-z_2)(z_{\mu_2}\delta_{\nu_1\nu_2 } - z_{\nu_2}\delta_{\nu_1\mu_2 }]
  +\nonumber \\{1\over 2}{\partial\over{ \partial z_{\nu_1}}}[D(z_1-z_2)(z_{\nu_2}\delta_{\mu_1\mu_2 } - z_{\mu_2}\delta_{\mu_1\nu_2 }]  )
  \end{eqnarray}
At $T\neq 0$ the electric field correlators do not coincide with the magnetic ones, and there are   four form factors, $D_E,D_{1E}, D_H,D_{1H}$ .
 
 For correlators of electric fields $E_{i_1} $ , $E_{i_1} $  Eq(15) gives
 \begin{equation}
  \Phi^{ab}_{i_1 i_2} =\delta^{ab}[\delta_{i_1 i_2}(D_E+{1\over 2}D_{1E})+{\partial \over{\partial i_1..}}]
  \end{equation}
  In the convolution with $\vec b_{\perp}$ the derivative terms give $0$. For the same reason
    
  $\delta_{i_1 i_2}   \to    \delta _{i_1 i_2} - {{k_{i_1}k_{i_2}} \over {k^2}}$

Going to the Fourier transform we get for   $\rho^a$ Eq(13)
 \begin{eqnarray}
  \rho^a= -{{q^2}\over 16}{\partial \over {\partial \beta}}[ \beta \int {{d^3k} \over {(2\pi)^3}} b^{i_1}_{\perp}(\vec k) b^{i_2}_{\perp}(-\vec k) \nonumber  \\D_E(k^2){1\over {k^2}}(k^2 \delta_{i_1 i_2}- k_{i_1} k_{i_2})]
\end{eqnarray}

 Here $\bar D_E (k^2)$ is the Fourier transform of $(D_E+{1\over 2}D_{1E})$.
 
 Since
  \begin{equation}
  (k^2 \delta_{ij} -k_{i1} k_{i2})b^{i_1}_{\perp}(\vec k)b^{i_2}_{\perp}(-\vec k) = |\vec H(\vec k)|^2
  \end{equation} 
  we can use  the explicit form of  $\vec H(\vec k)$
  \begin{equation}
  \vec H(\vec k)= \vec k \wedge \vec b_{\perp}(\vec k)
  \end{equation}
 with $\vec n$ the direction of the Dirac string ( we shall call it  $z$), and get
  \begin{equation}
   |\vec H(\vec k)|^2 = -{1\over {k^2}}+ {1\over {k^2_z}}
   \end{equation}
For  $\rho^a$ we then have
   \begin{equation}
   \rho^a =  {{q^2}\over 16}{\partial \over {\partial \beta}}[ \beta \int {{d^3k} \over {(2\pi)^3}}({1\over {k^2}}- {1\over {k^2_z}}) f(k^2)]
   \end{equation}
   where $f(k^2) \equiv {1\over {k^2}}D_E (k^2)$.
   
     Identifying our correlators with those of the stochastic model simply explains  that $\rho^a$ is independent both on $a$ and on the abelian projection.
   At large $\beta$(deconfined phase) $f(k^2)$ can be approximated by first order perturbation
   theory 
   \begin{equation}
   f(k^2) = {1\over {2Nk}}
   \end{equation}
  The only dependence on $\beta$ is the explicit factor in Eq(21) so that
  \begin{equation}
\rho^a=  {{q^2}\over {16N}} \int {{d^3k} \over {(2\pi)^3}}{1\over k}({1\over {k^2}}- {1\over {k^2_z}}) 
\end{equation} 
The integral is easily computed with UV cut-off ${1\over a}$ ($a$ the lattice spacing) and IR cut-off
 ${1\over {L_s a}}$  ($L_s$ the spacial  size of the lattice).
The result is
\begin{equation}
 \rho^a =  {{q^2}\over {16N}}{1\over {(2\pi)^2}}[-\sqrt(2) L_s + 2 ln(L_s) +const]
 \end{equation}
 By comparison with Eq(8) this  means that $\langle \mu^a \rangle = 0$ in the thermodynamical limit $L_s \to \infty$.
The correlators have been measured on the lattice both at $T=0$ and at $T \neq 0$. In the range of distances $.1fm \le  x \le 1fm$ they are well parametrized by a form \cite{10}
  \begin{equation}
  D_E = Ae^{-{x\over \lambda_b}} + {b\over {x^4}}e^{-{x\over \lambda_a}}
  \end{equation}
  with $\lambda_a\approx 2\lambda_b$ and $\lambda_b\approx .3fm$
  $A$ and $b$ are independent on $\beta$ within statistical errors in the range from $T=0$ up to $T\approx .95 T_c$ . Approaching further $T_c$ $A$ rapidly decreases  to zero.\cite{12}.
  
  In fact the  parametrization Eq(25) cannot be valid at shorter distances, where the operator product expansion and the non-existence of condensates of dimension less than 4 require that
  \begin{equation}
  D_E \approx {b\over 2} [{1\over {(x+ie)^4} }+{1\over {(x-ie)^4} }] + c + d x^2
  \end{equation}
  The prescription on the singularity is the same as in perturbation theory.
  At larger distances a stronger infrared cut-off at some distance  $\Lambda$, must exist,  since colored particles cannot propagate at infinite distance. This feature needs a further numerical investigation of correlators at large distance on the lattice.
  
  Up to  $T\approx .95 T_c$ the only dependence on $\beta$ in Eq(21) is again the explicit factor so that
  the result  coincides with Eq(24)  except that the lattice size $L_s$  is replaced by the infrared cut-off $\Lambda$
  \begin{equation}
   \rho^a =  {{q^2}\over 16}{1\over {(2\pi)^2}}[-\sqrt(2) \Lambda + 2 ln(\Lambda) +const]
   \end{equation}
   The integral of the exponential term of Eq(25) is included in the constant.
   This expression gives a finite value of $\rho^a$ for $T < T_c$ independent of the volume and hence $\langle \mu^a \rangle \neq 0$ .This  means dual superconductivity for any finite value of the UV cut-off $a$.  However in the continuum limit $a \to 0$ $\rho^a$ diverges so that $\langle \mu^a \rangle $ needs a renormalization.
   This is similar to what happens for the Polyakov line \cite{biel}. Existing Lattice data support this statement [See Fig(2) of ref.\cite{5}] ,but we plan to do a more systematic investigation of this issue.
   By approaching the critical temperature  both the IR cut-off $\Lambda$ and the coefficient $A$ in Eq(25) strongly depend on $\beta$ :$\Lambda $ diverges and $A $  tends to zero.
   A more detailed calculation, which will be reported elsewhere, gives
\begin{eqnarray}
   \rho^a = {{q^2}\over {16N}}{\partial \over {\partial \beta}}(\beta[{1\over {(2\pi)^2}}(-\sqrt(2){\Lambda\over a}+ 2\ln({\Lambda\over a}) \nonumber \\+const) +2N\lambda_b^4A_E(1+{\Lambda\over {\pi\lambda_b}}]
   \end{eqnarray}
   Numerical determinations  of the dependence of $A_E$ on of the temperature around $T_c$ exist in the literature\cite{12}. Not much is known about the behavior of $\Lambda$ . Further study is needed to understand how the scaling law
   of $\rho^a$ Eq(10) described in Sect 1 comes out  of  Eq(28).
   \section{Conclusions}
   We close with a few remarks.
   
   Checking the dependence $\rho^a \propto q^2$ is a test of the Stochastic Vacuum model.
   
   The independence of  $\rho^a$ on $a$ and on the abelian projection  are also an important test of it.
   
   The existence of confinement depends on a strong infrared cut-off of the field correlators. This can be directly checked on lattice..
    
    $\rho^a$diverges in the continuum limit, but provides a good description of confinement at any fixed value of the UV cut-off.
    
    An interesting interplay emerges of confinement  with infrared properties of gauge invariant field strength correlators.
    
    Useful discussions with H.G. Dosch, Y.A. Simonov,M.D'Elia, E. Meggiolaro, G.Paffuti are aknowledged.

\end{document}